\documentclass[11pt]{article}
\usepackage[top=1in, bottom=1in, left=1in, right=1in]{geometry}

\usepackage{amsmath}
\usepackage{parskip}
\usepackage{algorithm}
\usepackage{algorithmicx} 
\usepackage{algpseudocode} 
\usepackage{graphicx} 
\usepackage{amsthm}
\usepackage{physics}
\usepackage[round]{natbib}
\usepackage{xcolor}
\usepackage{booktabs}

\usepackage{url}
\usepackage{hyperref}
\usepackage{mathpazo}
\usepackage{subcaption}

\newcommand{\scl}[2][c]{%
  \begin{tabular}[#1]{@{}c@{}}#2\end{tabular}}

\title{On The Stability of Moral Preferences:\\ A Problem with Computational Elicitation Methods}

\usepackage{authblk}

\author[1]{Kyle Boerstler}
\author[2]{Vijay Keswani}
\author[2]{Lok Chan}
\author[2]{Jana Schaich Borg}
\author[3,4]{Vincent Conitzer}
\author[3]{Hoda Heidari}
\author[2]{Walter Sinnott-Armstrong}

\affil[1]{Activision}
\affil[2]{Duke University}
\affil[3]{Carnegie Mellon University}
\affil[4]{University of Oxford}

\date{}

\begin{document}

\maketitle

\begin{abstract}
Preference elicitation frameworks feature heavily in the research on participatory ethical AI tools and provide a viable mechanism to enquire and incorporate the moral values of various stakeholders. 
As part of the elicitation process, surveys about moral preferences, opinions, and judgments are typically administered only once to each participant. This methodological practice is reasonable if participants’ responses are \textit{stable} over time such that, all other relevant factors being held constant, their responses today will be the same as their responses to the same questions at a later time.  However, we do not know how often that is the case. It is possible that participants’ true moral preferences change, are subject to temporary moods or whims, or are influenced by environmental factors we don’t track.  If participants’ moral responses are unstable in such ways, it would raise important methodological and theoretical issues for how participants’ true moral preferences, opinions, and judgments can be ascertained. We address this possibility here by asking the same survey participants the same moral questions about which patient should receive a kidney when only one is available ten times in ten different sessions over two weeks, varying only presentation order across sessions. We measured how often participants gave different responses to simple (Study One) and more complicated (Study Two) controversial and uncontroversial 
repeated scenarios. 
On average, the fraction of times participants changed their responses to controversial scenarios (i.e., indicating instability) was around 10-18\%  ($\pm$ 14-15\%) across studies,
and this instability is observed to have positive associations with response time and decision-making difficulty. 
We discuss the implications of these results for the efficacy of common moral preference elicitation methods,  highlighting the role of response instability in potentially causing value misalignment between the stakeholders and AI tools trained on their moral judgments.
\end{abstract}

\section{Introduction}
The development of ethical and participatory AI tools in critical domains, such as healthcare or transportation, requires eliciting and modeling the moral judgments and values of various stakeholders \cite{borg2024moral}. 
To do so, a common methodology is to ask stakeholders to respond to various moral decision-making scenarios and use the resulting responses to learn task-specific moral preferences at an individual and population level.
However, a crucial assumption underlying this methodology is that participants' responses are \textit{stable}. 
Imagine that we want to design a healthcare resource allocation policy for a district and ask all the doctors in the district whether they prefer medical resources to be distributed equitably across all hospitals or distributed preferentially to maximize the number of treated patients.
Suppose a doctor expresses a preference to prioritize equity over efficiency one day, but on the very next day they say the opposite, even when nothing relevant has changed.
We would not know which policy the doctor really prefers, if either. 
This case illustrates a general phenomenon: the fact that a person expresses a preference on one occasion does not justify the inference that they will express the same preference on another occasion. Even if people's true underlying preferences are stable, any method that results in participant statements and choices changing over time in an unpredictable way is not a reliable method for eliciting, or ascertaining, stable participant preferences.  This lesson applies to any attempt to elicit preferences in surveys. It has been acknowledged in psychology literature through the expectation of reporting ``test-retest reliability'' measures for survey instruments \cite{aldridge2017assessing}.
Unfortunately, most preference elicitation surveys in AI fail to heed this warning. They ask participants a set of questions only once and cannot determine whether those participants would give different answers on different occasions due to changes in mood, decision strategy, or changes in opinion.

Many fields are affected by the issue of measurement instability, but here we will focus on impacts on the development of ethical and participatory AI tools.
AI is increasingly being used to make judgments in situations with moral consequences, such as when autonomous vehicles must determine which groups to prioritize when trying to avoid a collision \cite{awad2018moral} or to decide how to allocate scarce medical resources \cite{johnston2020preference,freedman2020adapting}. To endow AIs with the ability to make judgments in these situations in a way that is consistent with stakeholders’ values, AI researchers often try to learn stakeholders’ moral views using surveys that ask stakeholders to repeatedly choose among sets of moral alternatives, a process called ``\textit{moral preference elicitation}'' \cite{feffer2023preference}.  
Common preference elicitation procedures usually employ \textit{pairwise comparisons} between alternative actions, providing participants with an intuitive setup to express their preferences \cite{saaty2008relative}.
The AI's goals can then be constrained using models of stakeholders’ moral ``preferences'' learned through these elicitation procedures \cite{noothigattu2018voting}.  Critically, these procedures almost always administer surveys to stakeholders only once.    

Contrary to the assumptions of common moral preference elicitation procedures in AI ethics, moral psychology research shows that participants can change their moral judgments over time.
For instance, \citet{rehren2023stable} observe instability in moral judgments for scenarios involving \textit{sacrificial dilemmas}, where the participants are asked to choose between \textit{action} (sacrificing some people to save others) and \textit{inaction} (not acting and letting them die) -- e.g., the famous ``trolley problem''.
While this study provides evidence of instability in moral judgments, their applicability to AI-related settings is limited. 
This is because action-vs-inaction choice scenarios used in sacrificial dilemmas are structurally different than those used in AI-related preference elicitation literature.
Common preference elicitation frameworks use action-vs-action scenarios, i.e., scenarios that are posed as \textit{pairwise comparisons} between two actions
(e.g., \citet{srivastava2019mathematical, johnston2023deploying}).
Beyond structural differences, brain studies have shown that cognitive processes involved when making moral judgments in action-vs-action scenarios are different than those involved in action-vs-inaction scenarios \cite{schaich2006consequences}.
As such, it is unclear whether results from prior studies on instability for sacrificial dilemmas extend to preference elicitation frameworks commonly used in AI applications.

\paragraph{Our Contributions.}
	We explored this important methodological issue by asking survey participants to decide between moral options with the same relevant features, presented in different orders, on ten occasions over two weeks.  Our approach builds on the studies we described above by testing the stability of participants’ responses in a new moral context of medical triage decisions and asking participants to give responses across ten separate sessions, which allows us to better assess their response stability (or instability).  We also compared their response stability to uncontroversial ($>$90\% agreement across participants) and controversial ($<$75\% agreement across participants) decisions and investigated multiple hypotheses to determine the sources of instability in participant responses.
    The medical decisions we focus on pertain to kidney allocation.  When a kidney becomes available for transplant, it is often compatible with more than one needy recipient, and there are not enough donors (live or dead) to supply all patients in need.  As a result, doctors or hospitals often have to decide which one of several patients should receive a kidney that becomes available.  The patient who gets the kidney will typically gain decades of life by virtue of getting the transplant.  A patient who does not receive a kidney will have to keep waiting for a transplant, may lose quality of life as their condition deteriorates, and may even die waiting. Therefore, deciding who gets a kidney is a moral decision with life-and-death consequences.  

In our studies, we presented participants with pairs of kidney patients, who may differ across features, such as age, behaviors, and the number of dependents they have. Then, we asked participants to choose which of the two patients should be given the one available kidney. A subset of patient pairwise comparisons were repeated multiple times. Stability for a repeated comparison was measured by how often a participant chose the patient with the same features independently of changes in the order of the patients and features. 
Section~\ref{sec:study} reports and analyzes the degree of stability in participants’ responses to the repeated pairwise comparisons across sessions. 
Going beyond the repeated comparisons, Section~\ref{sec:response_models} studies the degree to which the statistical model that best fits patterns of all responses in one session continues to be the best fit for other sessions.
We find significant differences in predictions from models trained on different participant sessions, indicating that participants potentially change their decision-making model across sessions.

We also test multiple research questions to understand possible causes of lacks in participant response stability (Section~\ref{sec:hypotheses}). 
Specifically, we investigate whether response stability for any repeated pairwise comparison is associated with (1) attention/time taken to respond to the scenario, (2) the number of feature differences between the two patients in a repeated pairwise comparison, or (3) participant's perceived difficulty of making a judgment about the given comparison.
Our analysis provides evidence that response stability for any pairwise comparison is indeed associated with the response time and the {perceived difficulty} of the given comparison, with the evidence for the latter being most significant.
The results illuminate the scale and sources of response instability as well as whether and when common survey methods succeed in eliciting stable preferences. 

Note that our analysis primarily focuses on response stability at the participant level and not the population level; i.e., we assess each participant's response stability but we do not assess instability of aggregated judgments of all participants.
We discuss this point and the implications of instability on the efforts to develop AI tools that incorporate stakeholders' moral values in Section~\ref{sec:discussion}.  
Our instability results highlight the possibility of misalignment between stakeholders' moral values and the decisions of AI tools that utilize computational models of stakeholders' moral preferences.

\paragraph{Related Work.} 
Preference elicitation frameworks are employed in a wide variety of domains to learn people's preferences and incorporate them into downstream personalized applications \cite{jannach2005personalized}.
In the domain of moral judgments, \citet{awad2018moral} and \citet{noothigattu2018voting} study people's preferences over hypothetical moral dilemmas associated with the use of autonomous vehicles and ways to aggregate preferences obtained from a large population.
\citet{freedman2020adapting} employ elicitation frameworks to design tie-breaking mechanisms for kidney exchanges.
\citet{johnston2023deploying} propose using elicitation methods to understand people's preferences for healthcare resource allocation during crisis situations like COVID-19.
\citet{srivastava2019mathematical} use similar methods to model nuances associated with people's preferences over various mathematical notions of fairness.
Preference elicitation methods have also been employed for the development of participatory machine learning frameworks, where the goal is to actively seek out and incorporate diverse stakeholder opinions with the design of the automated system \cite{feffer2023preference, lee2019webuildai,evequoz2022diverse}.

Given this increasing popularity, it is important to simultaneously analyze their performance in practice.
Our work contributes to the recent literature investigating the effectiveness of computational methods for moral preference elicitation from the viewpoint of response stability.
In sacrificial dilemmas, \citet{rehren2023stable} found that 8-20\% of participants changed from saying that an agent should sacrifice some to save others to saying that the agent should not sacrifice some to save others in these dilemmas, or from the latter (“should not”) to the former (“should”). 
Their results align with previous findings on imperfect test-retest reliability when answering \textit{moral foundations questionnaire} \cite{curry2019mapping} and the impact of context on moral preferences \cite{schein2020importance}.
\citet{chan2020artificial} likewise found evidence of variability in participants' responses to kidney allocation judgments arising from the nature and source of feedback provided to them. 

Other works have raised other issues about preference elicitation efficacy. 
\citet{feffer2023moral} highlight concerns related to fairness and stability when aggregating moral preferences obtained from a heterogeneous population.
\citet{rogowski2024preferences} question the normative foundations underlying the use of preference elicitation for healthcare resource allocations, arguing that common aggregation methods may not necessarily lead to equitable or ``socially valuable'' outcomes.
Similar to moral dilemmas, \citet{conitzer2015crowdsourcing} discuss methods to model tradeoffs between various societal-level objectives and emphasize the importance of design considerations when asking for people's opinions on these tradeoffs and accounting for the heterogeneity of opinions among the relevant population.
While these works discuss important issues with applications of preference elicitation, our work questions the methodological assumptions of response stability when designing elicitation frameworks and highlights the impact of instability on ethical AI development.

\section{Study} \label{sec:study}

We performed two human subject experiments where we presented participants with a series of kidney allocation scenarios. 
Participants responded to these scenarios over a series of up to ten sessions spread out over two weeks. A subset of scenarios were repeated in each session to enable us to measure the stability of responses.

\subsection{Methods}

\paragraph{Participants.} We used Prolific
to gather our data for both studies with a 50/50 gender ratio (\textit{N}=30 in Study 1; \textit{N}=82 in Study 2). Exclusion criteria included: previous participation in a study from our group that involved kidney allocation queries, approval ratings $<$98\%, or completion of $<$50 previous submissions in Prolific.

\paragraph{Study Design.}
Participants were presented with scenarios that listed information about two fictional patients, each needing a kidney transplant (Figure 1). Participants were asked ``to choose which of two patients should receive the kidney when only one is available''.
In each scenario, each patient (named A or B) was described by the following features (set of possible feature values provided in parentheses):

\noindent
\textit{Study One:}
\begin{itemize}
    \item Decades of life that the patient is expected to gain from the transplant (0, 1, 2)
    \item Number of child dependents (0, 1, 2)
    \item Alcoholic drinks per day that the patient consumed prior to diagnosis (0, 2, 4)
    \item Number of past serious crimes committed (0, 1, 2)
\end{itemize}

\noindent
\textit{Study Two:}
\begin{itemize}
    \item Years of life expected to be gained from the transplant (0, 5, 10, 20, 25)
    \item Number of elderly dependents (0, 1, 2, 3)
    \item Years the patient had been on a waiting list for the transplant (1, 3, 5, 7)
    \item Hours a patient is expected to be able to work post-transplant (0, 10, 20, 30, 40, 50)
    \item Obesity (underweight, normal weight, overweight, obese, morbidly obese, very morbidly obese)
\end{itemize}

\begin{figure}
    \centering
    \includegraphics[width=0.5\linewidth]{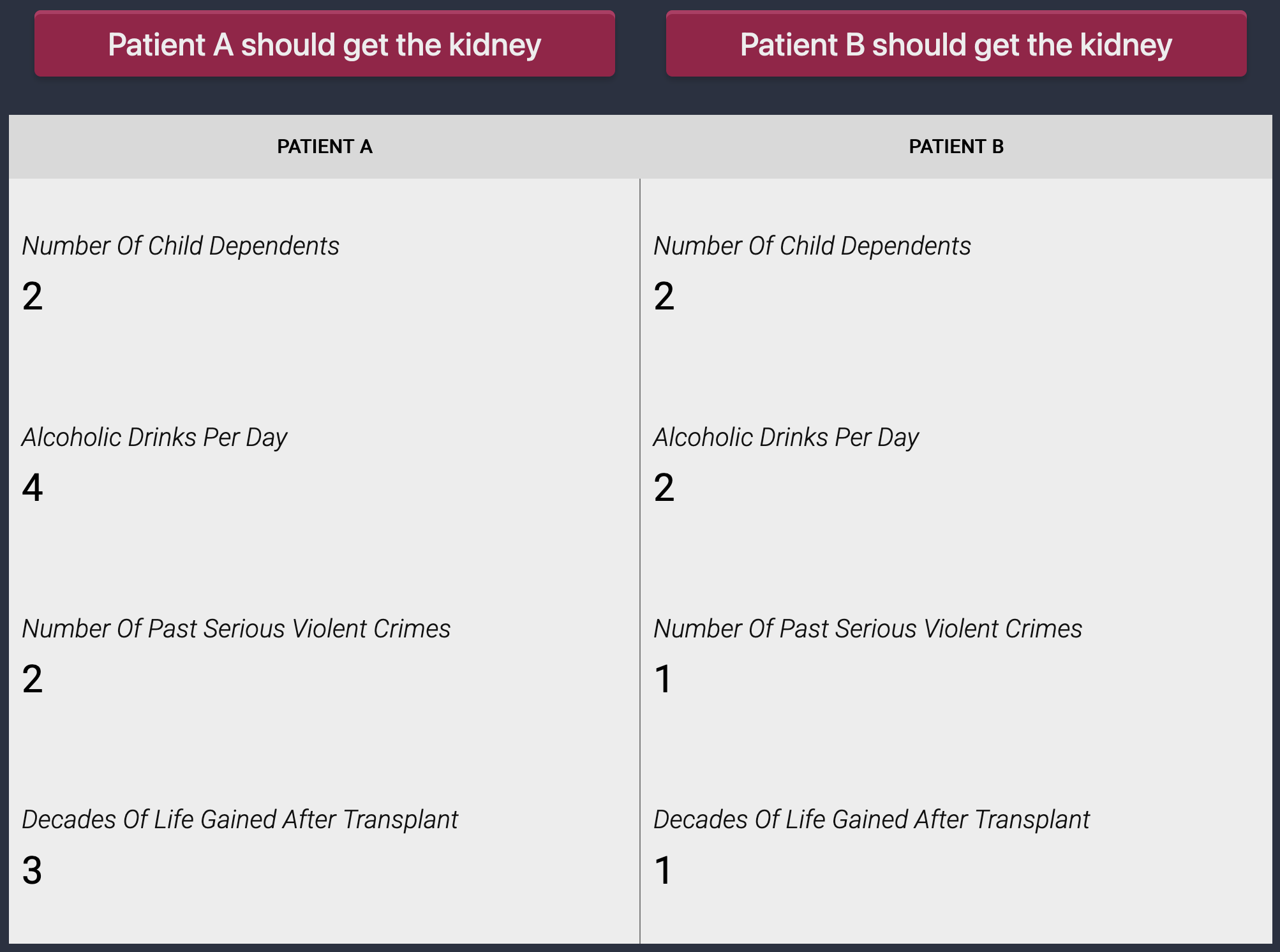}
    \caption{Example of the scenario interface participants responded to in Study One.}
    \label{fig:example}
\end{figure}

\noindent
 Participants’ choices in these scenarios are moral in nature since they affected harm to patients, were based on features (e.g., past crimes and alcohol abuse) that are often ascribed moral import, and responded to questions about who \textit{should} get the kidney (instead of whom they would give it to).

60 scenarios were presented to each participant during each session and participants took part in up to 10 sessions.
The ordering of the features top-to-bottom was randomized for each presented scenario (but kept the same for both patients in that scenario).  The majority of scenarios had values in each of the features for Patient A and B that were chosen randomly from the set of possible values ($N$ = 52 in Study One, $N$ = 46 in Study Two). 
Another 2 scenarios were attention checks where the life gained for one of the two patients was negative (e.g. –1 decade). Anyone paying attention would presumably not give a scarce kidney to a patient with this impossible loss of a decade of life. 

In each study, \textbf{six scenarios were repeated}. 
These repeated scenarios appeared once in each session for Study One and twice in each session for Study Two to test for stability within sessions. The values of features in these six repeated scenarios were chosen with the goal of making three repeated scenarios uncontroversial (so that almost all participants would agree on which patient should get the kidney; S1U1-S1U3 for Study One and S2U1-S2U3 for Study Two) and three of them controversial (so that participants would disagree about which patient should get the kidney; S1C1-S1C3 for Study One and S2C1-S2C3 for Study Two).  
Between sessions in Study One and between presentations within a session in Study Two, the patients in the repeated scenarios were randomly switched left to right and the features were randomly varied top to bottom to reduce the chance that participants would notice when scenarios were repeated and to encourage independent consideration of repeated scenarios. See Table~\ref{tab:scenarios} for the feature values in the initial presentation of these repeated scenarios for Study Two.  See Appendix Table~\ref{tab:scenarios_study_one} for the repeated comparisons used in Study One (S1U1-S1U3 and S1C1-S1C3).

\begin{table*}
    \centering
    \scriptsize
    \begin{tabular}{c|ccccc|cccc}
    \toprule
    &  \multicolumn{5}{|c|}{\textbf{Study Two} -- Scenario Feature Values} & \multicolumn{4}{|c}{\textbf{Study Two} -- Results} \\
        \scl{Repeated\\Scenarios} & \scl{Years\\Gained} & \scl{Elderly\\Dependents} & \scl{Years\\Waiting} & \scl{Work Hours\\After} & \scl{Obesity\\Level}&  &  \scl{Response\\Agreement } & \scl{Response\\Stability [std]} & \scl{Mean RT [std]/\\Median RT [IQR]} \\
    \midrule
        S2U1 & \scl{PA: 20\\PB: 5}  & \scl{PA: 2\\PB: 0}  & \scl{PA: 5\\PB: 3}  & \scl{PA: 30\\PB: 40}  & \scl{PA: N\\PB: OW} & & \scl{PA: 99.2\%\\PB: 0.8\%}  & \textbf{99.0\% [3.8\%]}  & 7.5 [8.3] / 5 [5] \\
    \cmidrule(lr){1-10}
        S2U2 & \scl{PA: 20\\PB: 10}  & \scl{PA: 1\\PB: 0}  & \scl{PA: 5\\PB: 5}  & \scl{PA: 30\\PB: 0}  & \scl{PA: OW\\PB: M} & & \scl{PA: 99.7\%\\PB: 0.3\%}  & \textbf{99.6\% [1.7\%]} & 7.5 [8.6] / 5 [5] \\
    \cmidrule(lr){1-10}
        S2U3 & \scl{PA: 10\\PB: 10}  & \scl{PA: 0\\PB: 0}  & \scl{PA: 1\\PB: 7}  & \scl{PA: 20\\PB: 30}  & \scl{PA: M\\PB: N} & & \scl{PA: 0.5\%\\PB: 99.5\%}  & \textbf{99.4\% [1.9\%]} & 7.8 [7.5] / 6 [5] \\
    \midrule
        S2C1 & \scl{PA: 25\\PB: 20}  & \scl{PA: 2\\PB: 1}  & \scl{PA: 5\\PB: 7}  & \scl{PA: 10\\PB: 30}  & \scl{PA: N\\PB: OW}&  & \scl{PA: 63.2\%\\PB: 36.8\%}  & \textbf{85.8\% [14.1\%]} & 9.8 [10.2] / 7 [7] \\
    \cmidrule(lr){1-10}
        S2C2 & \scl{PA: 10\\PB: 20}  & \scl{PA: 1\\PB: 0}  & \scl{PA: 5\\PB: 3}  & \scl{PA: 20\\PB: 40}  & \scl{PA: U\\PB: O} & & \scl{PA: 52.9\%\\PB: 47.1\%}  & \textbf{86.4\% [14.4\%]} & 9.6 [9.7] / 7 [7] \\
    \cmidrule(lr){1-10}
        S2C3 & \scl{PA: 5\\PB: 10}  & \scl{PA: 0\\PB: 2}  & \scl{PA: 7\\PB: 3}  & \scl{PA: 20\\PB: 30}  & \scl{PA: N\\PB:  O} & & \scl{PA: 71.9\%\\PB: 28.1\%}  & \textbf{87.9\% [14.5\%]} & 8.8 [8.1] / 6 [6] \\
    \bottomrule
    \end{tabular}
    \caption{Average between-participant agreement, average within-participant response stability [standard deviation], and average [standard deviation] and median [interquartile range] reaction times for Study Two (S2) uncontroversial (U) and controversial (C) repeated scenarios. For the Obesity feature, U = underweight, N = normal weight, OW = overweight, O = obese, M = morbidly obese, or V = very morbidly obese).}
    \label{tab:scenarios}
\end{table*}

\paragraph{Experimental Procedure.}
We asked our participants to respond to the 60 scenarios (with the composition described above) 
in each of 10 sessions on 5 days each week for 2 weeks. Not all participants completed all sessions. We excluded participants who contributed $<$300 responses (half of the scenarios requested). We also excluded all responses in any session in which a participant failed an attention check. 
In addition, after a participant selected a patient in Study Two, a window would pop up that said, “You are about to give the kidney to Patient A/B”, and the participant could not proceed without selecting a button with “Yes, continue” or another button with “No, go back.” We tracked how often participants went back to measure how often they saw their previous choice as a mistake. Our participants went back to change their answers in $<$2\% of our trials for any of our repeated scenarios, including the controversial scenarios, so we did not enter this factor into our analyses.

\subsubsection{Statistics and Analysis}

\paragraph{Outlier Removal.}  Participant's query reaction times (seconds) were right-skewed, so we excluded from all analyses query responses associated with reaction times that were 3 standard deviations beyond the mean, following a standard procedure for outlier removal \cite{berger2021comparison}.  

\paragraph{Response stability.}  To assess response stability, for each repeated scenario, we labeled the feature combination on the left half of the screen during the scenario’s initial presentation as “Patient A” (even if it was presented on the right in a later session), and the ones on the other half as “Patient B”. 
Stability was defined as the number of times a participant chose the patient that the participant chose more often, divided by the total number of times the scenario was repeated for that participant.  This means response stability can have a minimum value of 50\% and a maximum value of 100\%. For example, a participant who chose one patient (A or B) 8 times in a scenario repeated 10 times would be 80\% stable. 

\paragraph{Between-participant response agreement.}  We measured how often either Patient A or Patient B was chosen in aggregate by all participants. 
Uncontroversial scenarios were defined as those with $>$75\% agreement, and controversial scenarios were defined as those with $<$75\% agreement.

\paragraph{Modeling.} Different participants can use different decision-making processes to make kidney allocation decisions. To model each participant's process, e used the Bradley-Terry (BT) model to estimate the priority participants placed on patient features when making their allocation judgments \cite{bradley1952rank,hunter2004mm,freedman2020adapting}.  We assume that the effect each feature has on different patients across scenarios is the same. For example, the coefficient for alcohol for Patient A is the same as the coefficient for alcohol for Patient B. The estimation of the log(difference) for each patient winning in the BT model between two patients is modeled by:
\begin{align*}
    &\log \frac{Prob(A: Wins)}{Prob(B: Wins)} = \text{logit}(Prob(A: Wins)) \\
    &=  \left(\beta_{alco} \cdot p^A_{alco} + \beta_{dep} \cdot p^A_{dep} + \beta_{life} \cdot p^A_{life} + \beta_{crim} \cdot p^A_{crim}\right) \\
    &- \left(\beta_{alco} \cdot p^B_{alco} + \beta_{dep} \cdot p^B_{dep} + \beta_{life} \cdot p^B_{life} + \beta_{crim} \cdot p^B_{crim}\right)
\end{align*}
\noindent
This model can be further simplified to:
\begin{align}
     \text{logit}(Prob(A&: Wins)) = \beta_{alco} \cdot \text{diff}_{alco} + \beta_{dep} \cdot \text{diff}_{dep} \nonumber \\ &+ \beta_{life} \cdot \text{diff}_{life} + \beta_{crim} \cdot \text{diff}_{crim} \label{eq:bt},
\end{align}
where $\text{diff}_j = p^A_j{-}p^B_j$ (i.e., difference in value of feature $j$).

For the analysis in this section, we modeled features using their \textit{relative} values (i.e., difference in value of features of Patient A vs Patient B).
Section~\ref{sec:response_models} uses a broader setup, considering both raw and relative values of features.

\paragraph{Statistical Significance Tests.} The distributions of response stability values and reaction times were non-normal even after transformations, so we used Mann-Whitney U tests for significance testing, unless mentioned otherwise.

\subsection{Results}

17 participants (57\%) of our 30 recruited participants answered all 600 of the requested responses in Study One, and 19 answered 300 or more responses.  For Study Two, 29 participants of the 82 recruited participants (35\%) answered all 600 of our requested responses, and 52 responded to 300 or more responses. All results are based on the subsets of participants who answered 300 or more responses, $N$=19 for Study One and $N$=52 for Study Two.

%
Table~\ref{tab:scenarios} reports the between-participant agreement, within-participant response stability, and response time results for the six repeated scenarios in Study Two; Appendix Table~\ref{tab:scenarios_study_one} reports the same data for Study Two. 
In both studies, the three repeated scenarios that were intended to be uncontroversial (S1U1-S1U3; S2U1-S2U3) received $>$99\% response agreement across participants.  Although all three of the repeated scenarios that were intended to be controversial in Study Two (S2C1-S2C3) elicited $<$75\% agreement, only one of the repeated scenarios that were intended to be controversial in Study One (S1C3) met this criterion.

For Study Two, most participants were $>$90\% stable in their responses to the uncontroversial and controversial repeated scenarios averaged together (see Appendix Figure~\ref{fig:average_stability_boxplot}), but individual participants varied in their levels of stability for the controversial repeated scenarios from 50\% to 100\% (see Appendix Table~\ref{tab:participants_stability_levels}).  Only 5 out of 52 participants were perfectly stable for all repeated scenarios (Table~\ref{tab:participants_stability_levels}). Responses to the uncontroversial repeated scenarios were significantly more stable than responses to the controversial repeated scenarios ($p<$0.0001), and variance in stability levels was significantly higher for the controversial scenarios as well (F-test $p<$0.0001). For Study One, the difference in response stability between uncontroversial and controversial repeated scenarios is positive and statistically significant ($p<$0.01, Mann-Whitney test). However, the magnitude of stability differences between uncontroversial and controversial scenarios were smaller (compared to Study Two).

\subsection{Instability Causes} \label{sec:hypotheses}

To provide insight into what might cause participants' response instability, we investigate three research questions.  
\begin{itemize}
    \item RQ1: \textit{Is instability caused by insufficient time and attention to the decision?}
    \item RQ2: \textit{Is instability caused by challenges associated with having to consider multiple feature differences at once?}
    \item RQ3: \textit{Is instability a result of decisions being ``difficult'' because participants place similar priorities on the patients, given their feature values?}
\end{itemize}
 We will focus on Study Two results since it had more participants, greater response variability, and greater response instability than Study One and the results from the two studies were similar, but will highlight results from Study One when they differ from those in Study Two.

\begin{table}
    \centering
    \small
    \begin{tabular}{cccl}
    \toprule
        \scl{Grouping} & \scl{Mean [Std]/\\Median [IQR]\\ Seconds} & \scl{Percent\\Average\\Stability} \\
    \midrule
        \multicolumn{3}{c}{Study One} \\
    \midrule
        All repeated comparisons & 7.9 [9.1] / 5 [5] & 96.1\% \\
        Comparisons w/ RT $\geq$ 3 Sec. & 8.5 [9.3] / 5 [5] & 95.9\% \\
        Comparisons w/ RT $\geq$ 6 Sec. & 14.5 [11.5] / 9 [9] & 97.8\% \\
        Comparisons w/ RT $\geq$ 9 Sec. & 19.9 [12.4] / 15 [15] & 97.6\% \\
    \midrule
        \multicolumn{3}{c}{Study Two} \\    
    \midrule
        All repeated comparisons & 8.5 [8.8] / 6 [5] & 93.1\% \\
        Comparisons w/ RT $\geq$ 3 Sec. & 9.1 [9.0] / 6 [6] & 93.0\% \\
        Comparisons w/ RT $\geq$ 6 Sec. & 12.7 [10.3] / 9 [7] & 92.8\% \\
        Comparisons w/ RT $\geq$ 9 Sec. & 17.5 [11.8] / 13 [9] & 92.1\% \\
    \bottomrule
    \end{tabular}
    \caption{Response time statistics for each grouping of responses to repeated scenarios, and their aggregated stability.}
    \label{tab:rt_groups}
\end{table}

\subsubsection{Time on Task and Stability}

To test RQ1, we assessed stability across different subsets of responses to the repeated scenarios, defined by response times (Study One $N$ = 1109; Study Two $N$ = 5544; Table~\ref{tab:rt_groups}).   
No significant differences were found between the groups 
(ANOVA test to evaluate differences in mean stability for Table~\ref{tab:rt_groups} groupings had $p=0.92$ for Study One and $p=0.31$ for Study Two),
suggesting a negative response to RQ1, at least at the aggregate level.

A more nuanced picture emerged when, as an exploratory analysis, we analyzed individual response times to the three repeated controversial scenarios in Study Two, which were the scenarios that elicited the most instability (Table~\ref{tab:rt_study_two}).
We divided the responses to each controversial scenario into two groups: \textit{stable responses} -- those where the participant's response matches their dominant choice for the given scenario, and \textit{unstable responses} -- those where the participant's response does not match their dominant choice for the given scenario.
For each controversial scenario, average response times were significantly longer for unstable responses than stable responses ($p<$.0001, Mann-Whitney U test).   

\begin{table}
    \centering
    \small
    \begin{tabular}{cccc}
    \toprule
        \scl{Scenario} & \scl{Mean (Median)\\Seconds on Task for\\Stable Responses} & \scl{Mean (Median)\\Seconds on Task for\\Unstable Responses} & \scl{Signi-\\ficance\\Value}\\
    \midrule
        S2C1 & 10.0 (6.0) & 11.4 (8.0) & $<$.0001 \\
        S2C2 & 9.1 (6.0) & 13.7 (8.0) & $<$.0001 \\
        S2C3 & 9.1 (6.0) & 12.4 (8.0) & $<$.0001 \\
    \bottomrule
    \end{tabular}
    \caption{Mean (and Median) times taken by participants to respond to the three controversial Study Two scenarios, separated by stable and unstable responses. All differences between means for Stable versus Unstable responses were tested using a Mann-Whitney U test and the p-values are reported in the Significance Value column.}
    \label{tab:rt_study_two}
\end{table}

Since participant-level analyses might provide greater statistical sensitivity, as a final exploratory analysis, we assessed whether there were significant correlations between individual participants' average response times for a specific scenario and response stability values for scenarios S1C1-S1C3 (Study One, $N$=57) or S2C1-S2C3 (Study Two, $N$=156).  No significant correlations were found in Study One (\textit{p}=0.49), but a Pearson correlation of $-0.16$ (\textit{p}=0.05) was found in Study Two (see Figure~\ref{fig:stability_vs_rt} in the Appendix for scatter plots). Thus, we found no evidence that response instability was correlated with spending too little time or attention on a scenario. In contrast, our results suggest that a future study with greater statistical power might find evidence that instability is actually associated with longer reaction times.  Such an association is consistent with the idea that scenario \textit{complexity} or \textit{difficulty} contributes to response instability, which we investigate in the next sections.

\subsubsection{Feature Differences and Stability}
RQ2 asked whether instability could be caused by challenges associated with having to consider multiple feature differences at once.  Intuitively, perhaps participants make more mistakes or are not able to be sure of their answer when they have to compare differing values in a lot of features simultaneously, taxing their working memory and cognitive load. 
To assess whether this idea could have merit, we examined the number of features that differed between patients in each of our repeated scenarios. 
Table~\ref{tab:feat_diffs_study_two} shows the results of this analysis. 

We do not have enough variance in the number of feature differences to make strong conclusions, but we did not find compelling evidence for a relationship between the total number of feature differences and response stability. 
In Study One, the average response stability of scenarios with differences in three features was actually lower than that of scenarios with differences in four features ($p$=0.02, ANOVA main effect).  In Study Two, although response stability was lower on average for scenarios with five feature differences than for scenarios with three or four feature differences ($p{<}$0.0001, ANOVA main effect); responses to S2U1 were significantly more stable than S2C1, S2C2, and S2C3 ($p<$0.0001 for each test), even though the same number of features differed between patients in all of these scenarios. Thus, at least in our data, the number of feature differences in a repeated scenario does not likely account for much variance in response stability.


\begin{table}
    \centering
    \begin{tabular}{cccc}
    \toprule
        \scl{Scenario} & \scl{Total Feature Differences} & Stability \\
    \midrule
        \multicolumn{3}{c}{Study One} \\
    \midrule
        S1U1 & 3 & 99.0\% \\
        S1U2 & 4 & 99.5\%  \\
        S1U3 & 4 & 99.0\%  \\
        S1C1 & 3 & 95.3\% \\
        S1C2 & 3 & 93.9\%  \\
        S1C3 & 3 & 89.7\%  \\
    \midrule
        \multicolumn{3}{c}{Study Two} \\
    \midrule        
        S2U1 & 5 & 99.0\% \\
        S2U2 & 4 & 99.6\%  \\
        S2U3 & 3 & 99.4\%  \\
        S2C1 & 5 & 85.8\% \\
        S2C2 & 5 & 86.4\%  \\
        S2C3 & 5 & 87.9\%  \\
    \bottomrule
    \end{tabular}
    \caption{Total features that differ between patients in each of the repeated scenarios from Study One and Study Two, and the observed average stability levels for these comparisons.
    }
    \label{tab:feat_diffs_study_two}
\end{table}

\begin{figure*}[t]
    \centering
    \includegraphics[width=\linewidth]{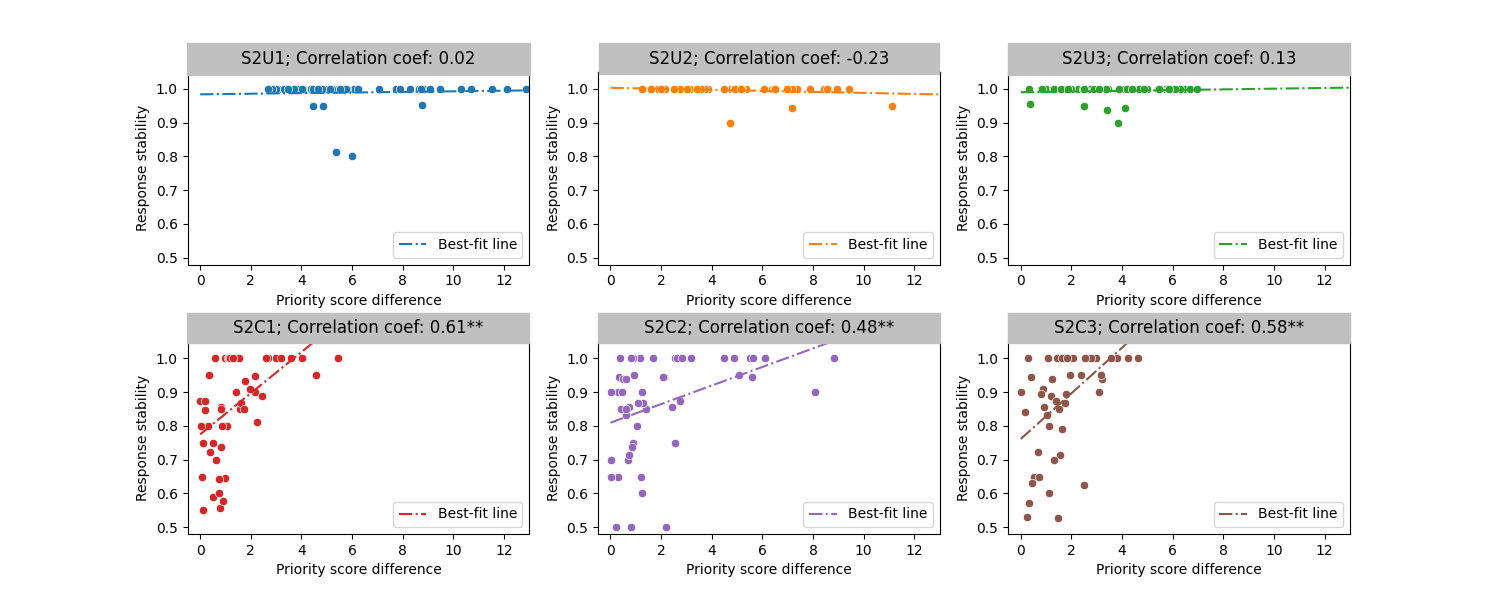}
    \caption{Scatter plots of response stability vs priority score difference for all six repeated scenarios in Study Two.
    Plot titles provide the Spearman correlation coefficient values; coefficients with the ``**'' mark indicate that the value is statistically significant at $p<$0.01.
    For the uncontroversial scenarios (S2U1-S2U3), most participants were perfectly stable. For the controversial scenarios (S2C1-S2C3), best-fit lines show a significant positive association between stability and priority score difference.}
    \label{fig:corr_study_two}
\end{figure*}

\subsubsection{Feature Weights and Stability}
RQ3 asked whether participants show instability when a decision is ``difficult'' because it seems like a close call.  
%
While prior work in psychology provides some evidence of an association between difficulty and uncertainty for \textit{value-based} decisions, it's unclear if this association holds in every setting as it can be abridged when participants are afforded enough time to ponder over difficult scenarios \cite{lee2021trading}.
In our setting, one way to think about this would be to assume that each participant assigns a certain importance or weight for each feature a patient has, and calculates (explicitly or implicitly) the ``priority score'' assigned to each patient by summing the weights of all the patients’ features (i.e., a linear weighted combination of patient features), and then chooses the patient with the highest priority score.  Scenarios with patients of similar priority scores could feel like close calls.

\begin{table}[t]
\small
\centering
\begin{tabular}{lc}
\toprule
 & \scl{\textit{Dependent Var:}\\ Stability} \\
\midrule
(Intercept)                               & $0.859^{***}$ \\
                                          & $(0.016)$     \\
Priority Score Difference                         & $0.022^{***}$ \\
                                          & $(0.003)$     \\
\midrule
Log Likelihood                            & $275.612$     \\
Num. obs.                                 & $312$         \\
Num. groups: ID                       & $52$          \\
Var: ID (Intercept)                   & $0.011$       \\
Var: ID $\times$ Priority Score Diff.             & $0.000$       \\
Cov: ID (Intercept) $\times$ Priority Score Diff. & $-0.002$ \\
Var: Residual                             & $0.007$       \\
\midrule
\multicolumn{2}{l}{\scriptsize{$^{***}p<0.001$; $^{**}p<0.01$; $^{*}p<0.05$}}
\end{tabular}
\caption{Mixed effects model for response stability vs priority score difference, with participant ID as group variable.}
\label{tab:coefficients}
\end{table}

To test this, we perform a Bradley-Terry (BT) analysis for each participant using their responses to all the pairwise comparisons specifically presented to them. Using BT analysis (and the regression subroutine in Eqn.~\eqref{eq:bt}), we first learn the relative weights assigned to all patient feature differences. The distributions of feature weights are visualized in Appendix Figures~\ref{fig:boxplot_study1} and \ref{fig:boxplot_study2}. We normalize each participant’s feature weight vector to have a quadratic norm of 1.
Note that Eqn.~\eqref{eq:bt} models features using differences in values across the two patients. One can also include raw feature values in this equation. However, including raw values does not lead to any improvement in the regression \textit{goodness-of-fit} measures (e.g., pseudo-$R^2$ values are mostly unchanged) in our case; hence, we mainly work with feature differences.

Next, we used the learned participant-level feature weights to quantify the approximate priorities that each participant assigned to each profile in a pairwise comparison. 
For instance, say $\beta^{(i)}$ denotes the vector containing the relative weights learned for participant $i$ using BT analysis. Suppose this participant is presented with a pairwise comparison $(p^A, p^B)$, where $p^A$ is a vector containing feature values for patient A and $p^B$ is a vector containing feature values for patient B. Then, \textbf{the difference in priority score assigned by participant $i$ for pairwise $(p^A, p^B)$ comparison} can be quantified as the absolute difference between the weighted sum of patient A and patient B profiles, i.e.,  
\[\textstyle \left| \sum_{j \in \mathcal{F}} \beta_j^{(i)} \cdot p_j^A - \sum_{j \in \mathcal{F}} \beta_j^{(i)} \cdot p_j^B\right|,\]
where $\mathcal{F}$ denotes the patient feature set, i.e., $\mathcal{F} =$\{number of elderly dependents, years of life gained, obesity level, weekly work hours, years on the waiting list\} for Study Two.

We test RQ3 by testing correlations between participants' learned priority score differences and response stability for all repeated scenarios. Results for Study One and Two are similar, but we focus on Study Two here due to its larger number of participants and a wider range of observed instability values.
Study One results are reported in 
Appendix~\ref{sec:study_one_appendix}.

Figure~\ref{fig:corr_study_two} presents the scatter plots of response stability vs priority score difference for all repeated scenarios in Study Two. For the uncontroversial scenarios S2U1, S2U2, and S2U3, most participants have perfect response stability and so the slope of the best-fit lines in these cases is almost 0. 
However, for the controversial scenarios S2C1, S2C2, and S2C3, we see significant positive associations between response stability and priority score difference (Pearson \textit{r}=0.43, $p{<}$0.01; Spearman $r_s$=0.54, $p{<}$0.01) for responses to controversial scenarios).

To assess whether individual participants drive the association between response stability and priority score difference, we fit a mixed-effects model for response stability, with priority score difference as the fixed effect and participant ID as the group variable with a random slope and intercept. Table~\ref{tab:coefficients} presents the results of this regression; even when accounting for variance across participants, the coefficient for priority score difference remains significant.

These analyses provide notable support for RQ3; i.e., the stability of a participant’s response for any pairwise comparison can track the closeness of the priority scores assigned to the two profiles by participants.

\section{Response Model Stability} \label{sec:response_models}

So far, we have limited our analyses of response stability to the 6 scenarios out of the 60 that were repeated in each session.  
While the repeated scenarios are useful in directly assessing whether participants provide different responses to the same scenario at different times, responses to non-repeated scenarios can also provide additional insight into the stability of participants' decision-making models across sessions.
To test this, we determined whether the statistical model that best fit each participant’s responses in one session was the same model that best fit the same participants’ responses in other sessions.
 
\subsection{Methods}

\subsubsection{Models}
We trained both linear and non-linear models on each participant’s data in each of their separate sessions, to cover a wide variety of hypothesis classes. 
Random Forests were chosen as the non-linear model for this analysis because they perform well off-the-shelf in many cases.
Since participants' responses are binary in this elicitation setting, we use Logistic Regression for the linear model.

All models take as input a description of any pairwise comparison between two patients $(p^A, p^B)$ and return a binary output indicating whether $p^A$ or $p^B$ should receive the kidney; in particular, the input to the model consists of the features of patient $p^A$, the features of patient $p^B$, and the difference vector between patient features, i.e. $p^A - p^B$.
We defined the best model for a participant in a session as the model among those tested that predicts the participant’s choices in that session with the greatest accuracy.

\subsubsection{Stability Evaluation}
We use each trained model to make predictions over the set of all possible pairwise comparisons (6,561 comparisons for Study One; 5,760,000 comparisons for Study Two).
To measure agreement between any two models, we calculated the fraction of times they agreed in their predictions to all presented comparisons.

We also check for robustness to model multiplicity (due to random state selection) by testing performance for different data splits. Appendix~\ref{sec:response_model_appendix} presents details of this analysis.
Overall, performance variation across data splits is minimal.

\subsubsection{Modeling Within-Participant Stability}
To measure stability within each participant across sessions on multiple days, we trained a different model on each separate session of each participant, made predictions with those models on the full set of all possible pairwise comparisons,
and compared the predictions of pairs of session-specific models. Two models agree for a pairwise comparison when they predict the same patient will be chosen to get the available kidney.
The average agreement between any two session-specific models (i.e., the fraction of comparisons where the two models have the same output) for a participant represents the stability of that participant across the two sessions.

\subsubsection{Modeling Between-Participant Consistency}
To measure consistency between participants, we trained a model on data from all sessions of participant \#1, repeated this procedure for participant \#2, and so on, to get a separate model for each participant. Then we made predictions with those models on the full set of all possible patient pairwise comparisons described above 
and calculated how much agreement there was between the models for each pair of participants.
This measure also serves as a baseline to compare within-participant stability against, since we expect consistency between participants to generally be lower than the response stability of each participant across their sessions.

\subsubsection{Baseline Comparison Model}
The best-fit model for a participant’s session may not be the true model the participant uses to make their judgments. It is also possible for two different models to make the same prediction; so, even if the best-fit models from two sessions make the same predictions, that does not mean a participant is using the same model in both sessions.  To generate a baseline comparison with these issues in mind, we determined how much models trained on subsets of data generated from the following decision policy would agree (each rule is applied in order and is ignored when values are equal across Patients A and B): 

\noindent
Study One:
\begin{itemize}
    \item If Patient A has more dependents than Patient B, give the kidney to Patient A (and vice-versa).
    \item If Patient A has less alcohol history than Patient B, then give the kidney to Patient A (and vice-versa).
    \item If Patient A has less criminal history than Patient B, then give the kidney to Patient A (and vice-versa).
    \item If Patient A has more life gained than Patient B, then give the kidney to Patient A (and vice-versa).
\end{itemize}

\noindent
Study Two:
\begin{itemize}
    \item If Patient A has been waiting longer than Patient B, give the kidney to Patient A (and vice-versa).
    \item If Patient A has more elderly dependents than Patient B, give the kidney to Patient A (and vice-versa).
    \item If Patient A will gain more life than Patient B, give the kidney to Patient A (and vice-versa).
    \item If Patient A has less obesity than Patient B, give the kidney to Patient A (and vice-versa).
    \item If Patient A will have more weekly work hours than Patient B, give the kidney to Patient A (and vice-versa).
\end{itemize}

\noindent
This model was applied to the same set of scenarios shown to each study participant to create a subset of hypothetical control judgments 
for each participant 
that represent the result of a policy that is applied with 100\% consistency.  We then trained linear and non-linear models on these sets of hypothetical control judgments and assessed their “stability” by computing the agreement of their judgments over the set of all possible pairwise comparisons
described earlier.

\begin{table*}
    \centering
    \small
    \begin{tabular}{cccc}
    \toprule
         Agreement & \scl{Between Real Participants\\(RF\% [LR\%])} & \scl{Within Real Participants\\(RF\% [LR\%])} & \scl{Within Baseline Condition\\(RF\% [LR\%])} \\
    \midrule
         Study One & \scl{High: 95.1 [96.7]\\Average: 75.2 [79.9] \\Low: 51.8 [52.3]} & \scl{High: 88.0 [89.6]\\Average: 80.9 [84.8] \\Low: 68.9 [71.5]} & \scl{High: 89.6 [91.7]\\Average: 88.3 [90.3] \\Low: 86.3 [89.0]} \\
    \midrule
         Study Two & \scl{High: 96.1 [96.5]\\Average: 73.4 [77.7] \\Low: 49.8 [53.4]} & \scl{High: 89.6 [90.0]\\Average: 79.6 [82.7] \\Low: 62.8 [67.2]} & \scl{High: 90.6 [90.9]\\Average: 88.6 [89.3] \\Low: 82.2 [83.5]} \\
    \bottomrule
    \end{tabular}
    \caption{Agreement levels among models in Studies One and Two.
    The between-participants analysis compares models trained on all sessions of different participants. The within-participants analysis (real and baseline) compares models trained on different sessions of the same participant. Random Forest and Logistic Regression results are reported as RF and LR respectively.}
    \label{tab:agreement_levels}
\end{table*}

\subsection{Results}
The average accuracy of random forests models for each participant was 91\% for Study One and 89\% for Study Two (see Appendix Tables~\ref{tab:model_perf_study_one}, \ref{tab:model_perf_study_two} for individual model results).

Table~\ref{tab:agreement_levels} reports the maximum, minimum, and average agreement between the best-fit models in the between-participant comparisons, within-participant comparisons, and the baseline condition.
For within-participant models, the average agreement using random forests was 80.9\% for Study One and 79.6\% for Study Two.
These values were significantly lower ($p<$ .0001 using the Mann-Whitney test) than the average stability of the baseline comparison model (88.3\% for Study One and 88.6\% for Study Two when using random forests), providing evidence that participants change their decision-making model across sessions.

For between-participant models, the average agreement using random forests was 75.2\% for Study One and 73.4\% for Study Two, as might be expected from the variety of weights individual participants placed on individual features (Figures~\ref{fig:boxplot_study1} and \ref{fig:boxplot_study2} show feature weight distributions from earlier BT analysis).  The agreement levels of between-participant models were smaller than those for within-participant models.  This suggests that even if individuals do seem to change their decision-making models across sessions, those models don't differ as much as what would be expected if they judged as a ``new person'' each session.

\section{Discussion} \label{sec:discussion}

\paragraph{Stability Comparisons.}
In Studies One and Two, we found 85-90\% stability—which is 10-15\% instability—in participants’ responses to every controversial scenario (S1C3 and S2C1-S2C3). In Section~\ref{sec:response_models}, a model-based analysis found an average of 79-81\% stability in the random forest models learned for different sessions of the same participant (average values in the second column of Table~\ref{tab:agreement_levels}). 
While the model-based analysis shows slightly lower stability values than those observed for the repeated comparisons, some level of disparity here should be expected since a learned model is unlikely to completely capture each participant's exact decision-making process.
Both analyses concur in the observation of a wide range of instability across participants.
Our data suggested that this instability might be at least partially explained by the difference between the total weights participants placed on the patient features in a scenario, which can be interpreted as how “close” the patients’ overall priorities were to each other for a given participant. 

When comparing results within the same participant to results between different participants from Section~\ref{sec:response_models}, we observe that stability within the same participant was only 5.7\% higher in Study One and 6.6\% higher in Study Two for non-linear models compared to stability between different participants (4.9\% higher in Study One and 5.0\% higher in Study Two for linear models). These small differences between within-participant model stability and between-participant model consistency are surprising if we expect people to agree with themselves much more than they agree with other people. Even so, these results line up well with previous studies \cite{rehren2023stable, curry2019mapping}.
Note that, both within participants and between participants stability levels are notably lower than the baseline condition.

The larger question raised by these results is whether and when common survey methods succeed in eliciting stable preferences. 
    %
    The instability we observe can be due to the sources we consider, due to weak or uncertain preferences, or perhaps, due to external circumstances (such as location or people around the participant).
    %
 %
This does not mean that common moral preference elicitation approaches are unsuitable for all circumstances.
In circumstances where 80-90\% average stability over time is enough, common surveys can be adequate without repeating the same questions at different times with the same subjects.
Our results also suggest that people might be stable in their responses to pairwise comparisons where the answer is very clear (i.e., large difference in the priorities assigned to each patient). 
Hence, if there is good reason to expect such clear preferences, then single-session approaches might be sufficient to learn moral preferences.
However, when more certainty is required, so that 80-90\% stability over time is not enough, and certain questions are expected to be difficult, researchers need to repeat the same questions at different times with the same subjects to determine whether their responses are stable. For example, if we are deciding whether to purchase a candy bar or a piece of fruit, it might not be so bad to waver and make different choices on different days. In contrast, if we are deciding on an issue of grave importance, such as who gets a kidney or some other critical but scarce resource, then more care should be taken to ensure that such a resource is not being allocated based on a temporary mood or whim rather than a stable preference or desire. Our findings suggest that policies about such important issues should not be based on studies that do not test for changes in responses over time.

\paragraph{Implications for Ethical AI Training.}
As mentioned earlier, a variety of recent studies employ elicitation frameworks to model people's moral judgments and use learned models to develop AI systems that incorporate stakeholders' moral values \cite{johnston2020preference, feffer2023preference, freedman2020adapting,srivastava2019mathematical}.
Response instability poses a serious challenge to this development pipeline as it can lead to inaccuracies in learned moral preferences.
Misalignment between the stakeholder's moral values and those learned by the AI system has the potential to cause significant harm when these tools are used in downstream applications.

Additionally, different sources of instability can have different kinds of impacts on AI systems that use these elicitation frameworks for training \cite{mcelfresh2021indecision}. 
Consider RQ3, which suggests that participants show instability when a decision is difficult (for which we found significant evidence). 
Here, the difficulty is characterized by the absolute difference between priority scores assigned to each patient in a pairwise comparison.
Yet, note that responses to difficult scenarios are the most informative about a participant's \textit{decision boundary}. 
Any attempt to accurately learn a participant's decision-making strategy will require modeling their decision boundary and will, correspondingly, require their responses to difficult scenarios.
The presence of increased response instability for difficult scenarios, therefore, poses a significant challenge to this methodology.

However, it is also possible for response instability to be part of the decision-making strategy for some participants.
For instance, in the case of RQ3, another explanation for instability in scenarios where the participant assigns similar priorities to the two patients is that the participant is simply \textit{indifferent} between two patients whose priority scores are close to each other.
In this case, they can choose different patients at different times; yet, their underlying decision strategy can still be considered stable/consistent.
However, in moral settings like kidney allocation, even though the decision-maker might be indifferent, the people affected by the decision (the kidney patients in our case) will not. 
Hence, even in these cases, developing ethical AI tools to assist in these moral settings requires computationally modeling when a participant might be indifferent and the resulting impact on response stability 
\cite{mcelfresh2021indecision}.

\paragraph{Preference Aggregation.}
Our analysis primarily operates at the participant level. 
We reported how often participants changed their responses to repeated scenarios and the degree to which session-specific models for each participant differed from each other.
One way to \textit{smooth over} the instability arising from each individual participant is to aggregate responses from all participants, as usually done in \textit{crowdsourcing} literature \cite{awad2020crowdsourcing, jiang2021can}.
However, aggregation may not be appropriate for personalized decision-aid tools tuned to the moral preferences of the people they are aiding. 
Additionally, aggregation comes with issues that can be exacerbated by response instability.

For instance, Table~\ref{tab:scenarios} reports the aggregated measure ``Response agreement'', quantifying the percent times each patient in the repeated scenarios was chosen. Notice that, for S2C2, Patient A was chosen in 52.9\% of the cases. However, response stability in this case was 86.4\%.
Hence, while patient A might seem like the dominant choice when all responses are aggregated, the observed average instability of 14.6\% makes it difficult to be certain of this choice.

Beyond final decision uncertainty, preference aggregation in moral decision-making settings can also dilute minority opinions when the underlying population is sufficiently heterogeneous \cite{feffer2023moral}.
To assess decision heterogeneity across participants, certain measures in crowdsourcing literature allow one to test for ``inter-rater reliability'', i.e., degree of consistency across participants, using measures like Cohen's $\kappa$ \cite{mchugh2012interrater} or Krippendorff's $\alpha$ \cite{krippendorff2018content}.
However, \textit{intra-rater reliability}, using measures like response stability, has received little attention and our paper goes toward filling this gap for moral preference elicitation frameworks.

\paragraph{Limitations and Future Directions.}
These studies were limited in several ways. Our samples were small and concentrated in the United States, so they are not representative or diverse enough for generalizations. 
Our participant population consists of people who don't have any immediate stakes associated with their decisions. If the same surveys are presented to doctors and/or kidney patients, we might get different sources and scales of instability.
Similarly, we presented our participants with only one kind of choice: kidney allocations. 
We might get very different results for other kinds of moral choices
in AI domains, e.g., autonomous vehicles \cite{awad2018moral} or policy development \cite{johnston2023deploying}, as well as for frameworks beyond pairwise comparisons, e.g., when participants rank choices \cite{ali2012ordinal} or when they provide motivations along with their choices \cite{siebert2022estimating}.
In terms of modeling, while Bradley-Terry analysis is a standard approach in the pairwise comparison setting, it does make certain assumptions (e.g., acyclic preferences) whose impact on instability analysis can also be assessed in future work on this topic.

Our surveys did not reveal definitive evidence for any sole mechanism or cause of instability over time. We considered time-on-task, feature differences, and feature weights as possible explanations for our results.  While our findings were in some cases indicative of their role in instability, we do not have causal evidence. Future studies should manipulate these and other potential explanations to determine the mechanisms that produce instability.
Additionally, methods to handle response instability, either through explicit modeling of decision mechanisms and strategies that may contribute to instability---such as indecision or indifference when option priorities are close  \cite{mcelfresh2021indecision}---or by asking participants to provide multiple responses/explain their responses \cite{liscio2024value,siebert2022estimating}---can also be explored as part of future work.

Despite these limitations, these studies have highlighted a vulnerability in AI development practices that assume participants’ responses to elicitation surveys are stable. These results suggest caution when training ethical AI systems based on moral queries that are only asked once.
Incorporating stakeholders' moral values within AI systems is a formidable task and
challenges like response instability make it difficult for AI developers to effectively elicit and employ people's moral judgments. 
Nevertheless, addressing these challenges can help develop robust AI tools that abide by the moral values of the relevant stakeholders.

\section*{Acknowledgements}

We thank Ya-Yun Huang for help with the BT analyses. We are grateful for the financial support from OpenAI and Duke.

\clearpage
\bibliography{references}

\appendix
\clearpage

\section{Additional Results for Stability Analysis} \label{sec:study_one_appendix}

In this section, we present results that were excluded from the main body.

Table~\ref{tab:scenarios_study_one} presents the repeated comparisons and corresponding stability results for Study One. Stability levels are again lower for the scenarios that were intended to be controversial (S1C1-S1C3) compared to scenarios intended to be controversial (S1U1-S1U3). However, average stability was less than 90\% for only one scenario (S1C3).

We also report additional results that were relevant to the research questions posed in Section~\ref{sec:hypotheses}.
For RQ1, we test associations between response time and stability for the repeated scenarios. 
Scatter plots (along with best-fit lines) between these two variables are presented in Figure~\ref{fig:stability_vs_rt}.
For Study One, the correlation between response time and stability is not statistically significant.
For Study Two, this correlation is negative (-0.16) and statistically significant, providing some evidence of a possible negative association between response time and stability.

For RQ3, we test associations between priority score difference and response stability. 
In this analysis, we first compute the relative weights assigned to all patient features by each participant.  
Figures~\ref{fig:boxplot_study1} and \ref{fig:boxplot_study2} present the weight distributions learned for Study One and Study Two respectively.
We also primarily presented results regarding RQ3 for Study Two in the main body. The results for Study One are similar. In particular, the scatter plots between priority score difference and response stability for all Study One repeated scenarios are presented in Figure~\ref{fig:scatter_by_pairid_study1}.
Once again, there is no significant association between priority score difference and stability for the uncontroversial scenarios (S1U1-S1U3). However, for S1C1-S1C3, there is again a positive association between priority score difference and stability, indicating that participants were more unstable for scenarios that they found difficult. 

\section{Additional Results for Response Model Analysis} \label{sec:response_model_appendix}
The accuracies of random forest models trained on each participant's data for Study One and Study Two are presented in Tables~\ref{tab:model_perf_study_one} and \ref{tab:model_perf_study_two} respectively.
Details of additional robustness checks related to model multiplicity are provided below.

\paragraph{Model Multiplicity. }
Training a Machine Learning model on the same data while using two different random states as initializations will sometimes result in two drastically different models. Depending on the data, two models may agree significantly in the center of the distribution but then disagree wildly further from the center of the distribution. If this problem of model multiplicity infects our analyses, then our results regarding agreement between models might be mere coincidence based on the random initialization of the models. This problem might arise even if we choose the same random state across participants’ data (which we did) because that random state could cause drastic variance in the resulting model for a given subset of participants and not for the others.

	To show that our analyses do not fall prey to this problem, we used Sci-Kit Learn’s implementation of Stratified Shuffle Split and shuffled the data of each participant 30 times, training a model for each and evaluating the test portion of each shuffle. We found the best-performing model based on these 30 splits, compared models from the shuffling that were 4-6\% less accurate on the test set, and then tested for agreement in the same way described above for our between-participant agreement analysis.
	The results of this analysis show a range from 90-98\% agreement between the best and least accurate models of each participant (both linear and non-linear). This suggests that our analyses do not suffer from the problem of model multiplicity.

\begin{figure}
    \centering
    \includegraphics[width=0.5\linewidth]{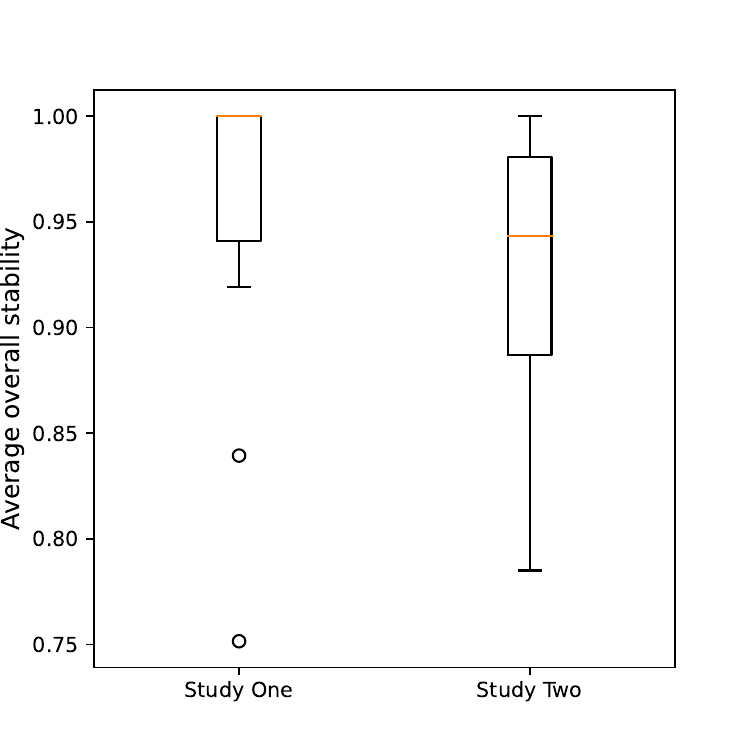}
    \caption{Distribution of average stability levels across participants for the two studies.}
    \label{fig:average_stability_boxplot}
\end{figure}

\begin{table}
    \centering
    \begin{tabular}{ccccccc}
    \toprule
        & \multicolumn{6}{c}{Stability levels}\\
        Scenario & 50\% & 60\% & 70\% & 80\% & 90\% & 100\% \\
    \midrule
        S1C3    & 0 & 4	& 1	& 1	& 1	& 12 \\
        S2C1	& 4	& 5	& 6	& 13 & 7 & 20 \\
        S2C2	& 4	& 4	& 7	& 10	& 15	& 15 \\
        S2C3	& 3	& 5	& 3	& 10	& 11	& 23 \\
    \bottomrule
    \end{tabular}
    \caption{Number of participants with different levels of stability in controversial scenarios of Study One (S1C3) and Study Two (S2C1, S2C2, S2C3). Each column represents the number of participants that fall within that stability range up to the stability range in the next column. The 100\% column represents the number of participants that attained 100\% stability for the controversial scenario in that row.}
    \label{tab:participants_stability_levels}
\end{table}

\begin{table*}[bp]
    \centering
    \small
    \begin{tabular}{c|cccc|cccc}
    \toprule
    &  \multicolumn{4}{|c|}{\textbf{Study One} -- Scenario Features} & \multicolumn{4}{|c}{\textbf{Study One} -- Results} \\
        \scl{Repeated\\Scenarios} & \scl{Serious\\Crimes} & \scl{Child\\Dependents} & \scl{Alcoholic\\Drinks} & \scl{Decades\\Gained} &  &  \scl{Percent\\Agreement} & \scl{Response\\Stability} & \scl{Response\\Time} \\
    \midrule
        S1U1 & \scl{PA: 2\\PB: 0}  & \scl{PA: 0\\PB: 0}  & \scl{PA: 4\\PB: 0}  & \scl{PA: 1\\PB: 3}  & & \scl{PA: 1.0\%\\PB: 99.0\%}  & \textbf{99.0\% [2.8\%]}  & 7.5 [7.5] / 5 [5] \\
    \cmidrule(lr){1-9}
        S1U2 & \scl{PA: 0\\PB: 2}  & \scl{PA: 2\\PB: 0}  & \scl{PA: 0\\PB: 4}  & \scl{PA: 3\\PB: 1}  & & \scl{PA: 99.5\%\\PB: 0.5\%}  & \textbf{99.5\% [2.2\%]}  & 7.5 [9.8] / 4 [4] \\
    \cmidrule(lr){1-9}
        S1U3 & \scl{PA: 0\\PB: 2}  & \scl{PA: 1\\PB: 0}  & \scl{PA: 0\\PB: 4}  & \scl{PA: 3\\PB: 1}  & & \scl{PA: 99.0\%\\PB: 1.0\%}  & \textbf{98.9\% [3.1\%]}  & 7.8 [9.6] / 4 [5] \\
    \midrule
        S1C1 & \scl{PA: 2\\PB: 0}  & \scl{PA: 0\\PB: 2}  & \scl{PA: 0\\PB: 2}  & \scl{PA: 2\\PB: 2}  & & \scl{PA: 8.0\%\\PB: 92.0\%}  & \textbf{95.3\% [12.3\%]}  & 7.5 [8.4] / 4 [5] \\
    \cmidrule(lr){1-9}
        S1C2 & \scl{PA: 0\\PB: 1}  & \scl{PA: 1\\PB: 0}  & \scl{PA: 2\\PB: 0}  & \scl{PA: 2\\PB: 2}  & & \scl{PA: 92.1\%\\PB: 7.9\%}  & \textbf{93.9\% [13.4\%]}  & 8.0 [8.6] / 5 [5] \\
    \cmidrule(lr){1-9}
        S1C3 & \scl{PA: 0\\PB: 2}  & \scl{PA: 0\\PB: 1}  & \scl{PA: 2\\PB: 0}  & \scl{PA: 2\\PB: 2}  & & \scl{PA: 41.0\%\\PB: 59.0\%}  & \textbf{89.7\% [15.1\%]}  & 9.3 [10.1] / 5 [6] \\
    \bottomrule
    \end{tabular}
    \caption{Study One repeated comparisons and main results. The first part of the table presents features of uncontroversial (U) and controversial (C) repeated scenarios in Study One (S2). The second part of the table presents agreement, stability, and response times for the repeated scenarios. Stability is reported in mean [standard deviation]. Response times are reported in mean [standard deviation]/median [interquartile range] seconds.}
    \label{tab:scenarios_study_one}
\end{table*}

\begin{figure*}
\centering
\begin{subfigure}{.5\textwidth}
  \centering
  \includegraphics[width=\linewidth]{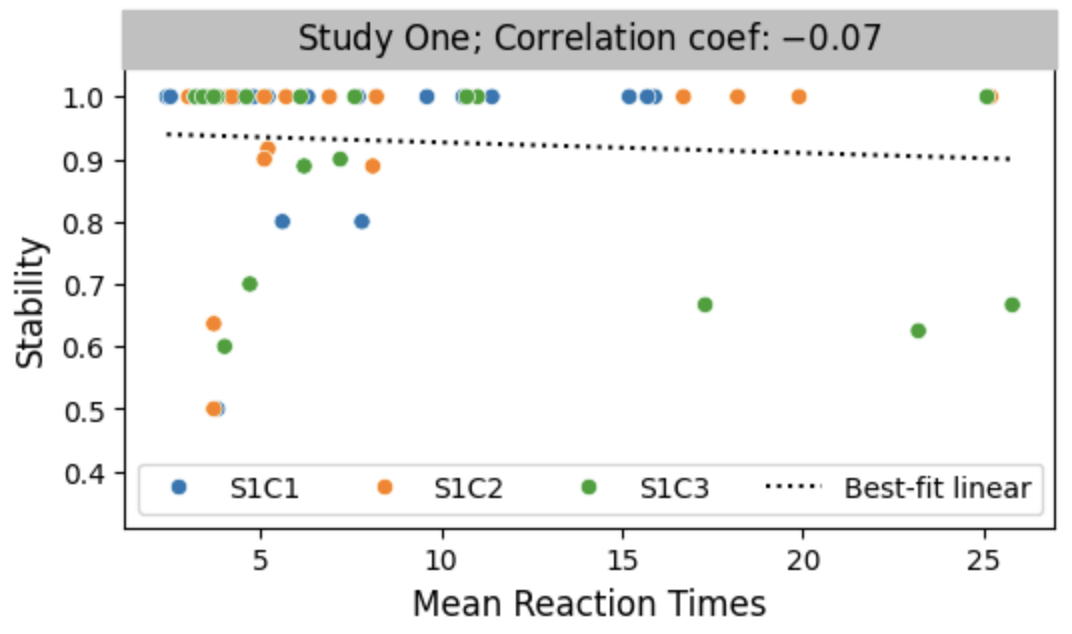}
  \caption{Study One}
\end{subfigure}%
\begin{subfigure}{.5\textwidth}
  \centering
  \includegraphics[width=\linewidth]{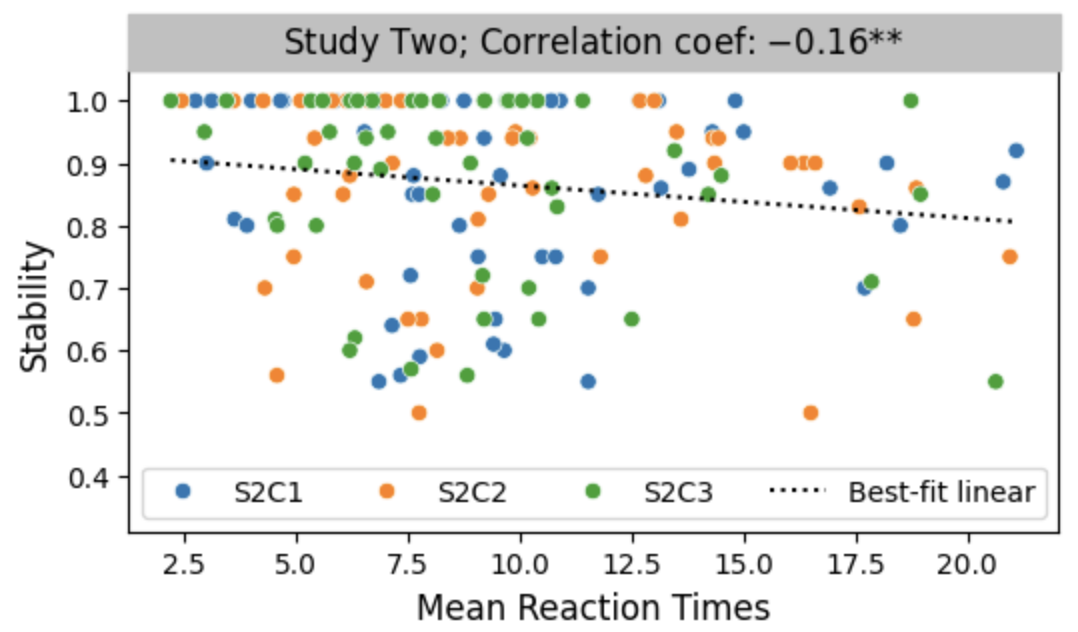}
  \caption{Study Two}
\end{subfigure}
\caption{Scatter plot of mean reaction time vs response stability for controversial repeated scenarios in Study One and Study Two. The plots also present the best-fit lines and Pearson correlation coefficient between these two variables (** indicates that the correlation was statistically significant at $p<0.05$).}
\label{fig:stability_vs_rt}
\end{figure*}

\begin{figure*}
    \centering
    \includegraphics[width=0.8\linewidth]{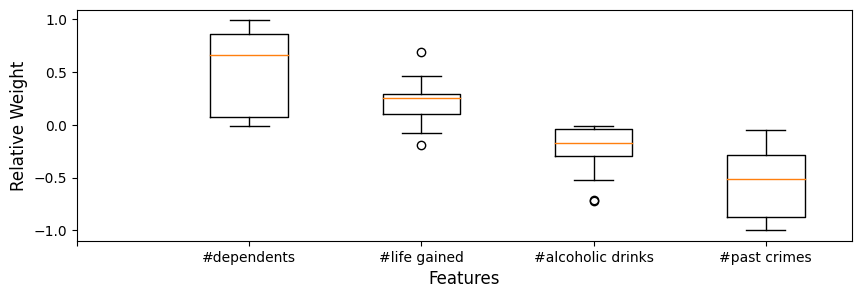}
    \caption{Summary statistics for feature weights learned from the Bradley-Terry model for each participant in Study One.}
    \label{fig:boxplot_study1}
\end{figure*}

\begin{figure*}
    \centering
    \includegraphics[width=0.8\linewidth]{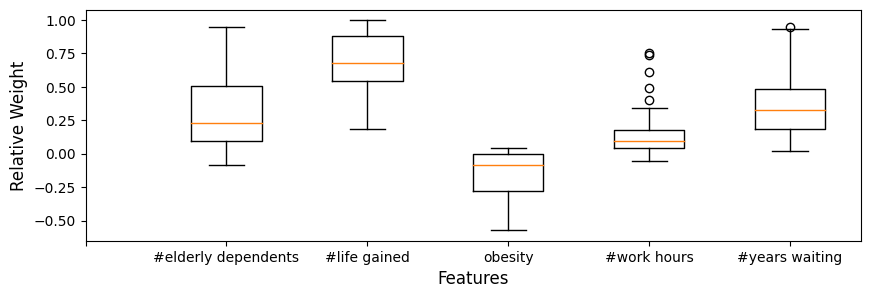}
    \caption{Summary statistics for feature weights learned from the Bradley-Terry model for each participant in Study Two.}
    \label{fig:boxplot_study2}
\end{figure*}

\begin{figure*}[t]
    \centering
    \includegraphics[width=\linewidth]{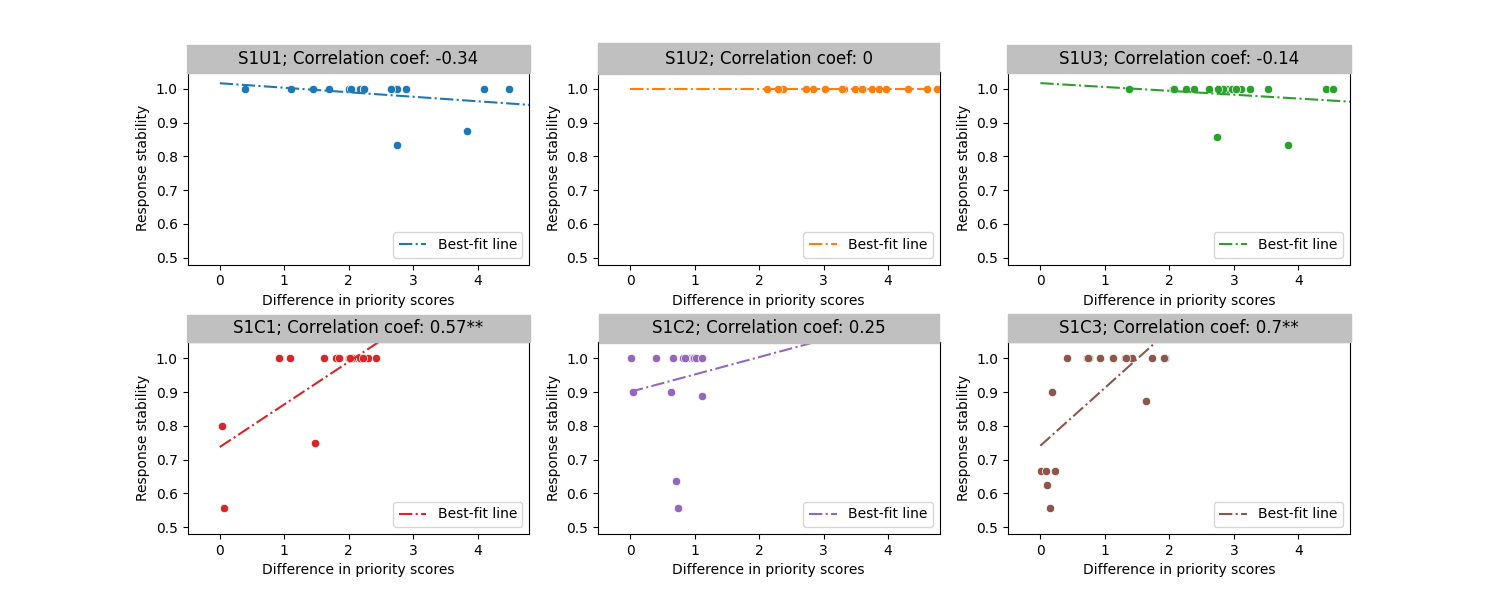}
    \caption{Scatter plots of stability vs priority score difference for all six repeated scenarios in Study One. For the uncontroversial scenarios (S2U1-S2U3), most participants were again perfectly stable. For the controversial scenarios (S1C1-S1C3), once again best-fit lines show a positive association between stability and priority score difference.}
    \label{fig:scatter_by_pairid_study1}
\end{figure*}

\begin{table}[]
    \centering
    \begin{tabular}{c|c}
    \toprule
        Participant &	Random Forest Accuracy\\
        \midrule
        2 & 93\% \\ 
        5 & 80\% \\ 
        6 & 97\% \\ 
        8 & 94\% \\ 
        9 & 93\% \\ 
        11 & 95\% \\ 
        12 & 95\% \\ 
        13 & 94\% \\ 
        14 & 85\% \\ 
        16 & 85\% \\ 
        17 & 93\% \\ 
        19 & 92\% \\ 
        23 & 79\% \\ 
        24 & 93\% \\ 
        25 & 96\% \\ 
        27 & 93\% \\ 
        28 & 93\% \\ 
        29 & 91\% \\ 
        30 & 89\% \\ 
        Average & 91\% \\ 
    \bottomrule
    \end{tabular}
    \caption{Model performance per participant in Study One.}
    \label{tab:model_perf_study_one}
\end{table}

\begin{table}[]
    \centering
    \footnotesize
    \begin{tabular}{c|c}
    \toprule
        Participant &	Random Forest Accuracy\\
        \midrule
        2 & 85\% \\ 
        5 & 85\% \\ 
        7 & 93\% \\ 
        9 & 87\% \\ 
        11 & 95\% \\ 
        14 & 91\% \\ 
        15 & 87\% \\ 
        16 & 86\% \\ 
        17 & 91\% \\ 
        18 & 92\% \\ 
        19 & 91\% \\ 
        20 & 89\% \\ 
        21 & 87\% \\ 
        23 & 88\% \\ 
        24 & 94\% \\ 
        26 & 91\% \\ 
        29 & 76\% \\ 
        31 & 91\% \\ 
        33 & 90\% \\ 
        36 & 93\% \\ 
        37 & 89\% \\ 
        38 & 82\% \\ 
        39 & 93\% \\ 
        41 & 74\% \\ 
        42 & 92\% \\ 
        43 & 90\% \\ 
        44 & 85\% \\ 
        46 & 93\% \\ 
        47 & 93\% \\ 
        49 & 86\% \\ 
        51 & 94\% \\ 
        53 & 90\% \\ 
        54 & 89\% \\ 
        55 & 96\% \\ 
        56 & 95\% \\ 
        57 & 89\% \\ 
        58 & 83\% \\ 
        60 & 79\% \\ 
        64 & 88\% \\ 
        65 & 92\% \\ 
        66 & 86\% \\ 
        67 & 93\% \\ 
        69 & 91\% \\ 
        70 & 91\% \\ 
        72 & 82\% \\ 
        73 & 95\% \\ 
        74 & 91\% \\ 
        75 & 83\% \\ 
        76 & 72\% \\ 
        77 & 88\% \\ 
        79 & 90\% \\ 
        80 & 84\% \\ 
        82 & 93\% \\ 
        Average & 89\% \\ 
    \bottomrule
    \end{tabular}
    \caption{Model performance per-participant in Study Two.}
    \label{tab:model_perf_study_two}
\end{table}

\end{document}